\documentstyle[prc,preprint,aps]{revtex} 
\tighten 
\begin{document} 
\draft 
\title{Neutron-proton pairing in the BCS approach} 
\author{O. Civitarese$^a$, M. Reboiro$^a$, and P. Vogel$^b$ }
\address{$^a$ Department of Physics, University of La Plata, C.C. 67,
La Plata, Argentina, \\
$^b$Department of Physics 161-33, Caltech,  Pasadena, CA 91125,\\ 
}
\date{\today} 
\maketitle

\begin{abstract}
We investigate the BCS treatment of neutron-proton pairing involving
time-reversed orbits. We conclude that an isospin-symmetric 
hamiltonian, treated with the help of the generalized 
Bogolyubov transformation, fails
to describe the ground state pairing properties correctly. 
In order for the $np$ 
isovector pairs to coexist with the like-particle pairs, 
one has to break the isospin symmetry
of the hamiltonian by artificially increasing 
the strength of $np$ pairing
interaction above its isospin symmetric value. We conjecture that the
$np$ isovector pairing represents part (or most) 
of the congruence energy
(Wigner term) in nuclear masses.
\end{abstract} 

\pacs{21.60.Fw, 23.40.-s, 23.40.Hc}

\section{Introduction}

Pairing correlations are an essential feature 
of nuclear structure \cite{BM}.
In proton-rich nuclei
with $N \approx Z$ the neutron and proton fermi levels are close 
to each other and therefore the neutron-proton
($np$) pairing correlations can be expected to play 
a significant role in their
structure and decay (for a review of the early work 
on $np$ pairing theory see
Ref.\ \cite{Goodman}). In contrast, in the heavier nuclei 
with large neutron excess 
the neutron-proton pairing correlations can be usually neglected.

There has been a recent revival of interest in the 
theoretical description of 
pairing involving both neutrons and 
protons \cite{Engel1,Engel2,Civitarese96,Wyss}. 
This renaissance stems from the advent of experiments 
with radioactive beams,
as well as from the application of neutron-proton 
pairing concepts in the description
of alpha decay \cite{Liotta} and double beta decay 
\cite{Cheoun,Pantis}.
However, the theoretical treatment is not without a controversy.
While intuition and arguments of isospin symmetry 
suggest that the neutron-proton
pairing correlations should be as important as the like particle 
pairing correlations in the $N \simeq Z$ nuclei, 
the balance between these pairing
modes is delicate and the standard approximations often fail.

In order to elucidate what is going on we examine 
the treatment of neutron-proton pairing in the generalized
Bogolyubov transformation approach, in particular 
the role of isospin symmetry.
The problem at hand is the determination of the ground state 
of an even-even system with the hamiltonian
\begin{equation}
H=\sum_{jmt}  \epsilon _{jt} a_{jmt}^{\dagger} a_{jmt} - 
     \frac 14  \sum_{jmj^{\prime }m^{\prime }} \sum_{tt^{\prime }} 
                           G_{ t t^{\prime } } a_{jmt}^{\dagger}
                            a_{\overline{jm}t^{\prime }}^{\dagger} 
                      a_{\overline{j^{\prime}m^{\prime }}t^{\prime }}
                           a_{j^{\prime }m^{\prime }t},
\label{e:ham}
\end{equation} 
where $(jmt)$ represents the angular momentum, its 
projection and the isospin projection of the single-particle (s.p.)
state created (annihilated) by the operator 
$a_{jmt}^{\dagger}$ $(a_{jmt})$, and as usual
$a_{\overline{jm}t} = (-1)^{j-m}a_{j-mt}$. The three
coupling constants $G_{t t^{\prime}}$
(we assume that $G_{t t^{\prime}}=G_{t^{\prime}t}$)
characterize the 
monopole pairing interaction. The interaction 
couples only states in 
time reversed orbits, but allows an arbitrary combination
of the isospin projection indeces. Obviously, when isospin symmetry
is imposed, the s.p. energies become independent 
of the isospin label $t$, and 
$G_{nn} = G_{pp} = G_{pn} = G_{np} \equiv G$.  
The interaction then
describes the isovector $T=1$ pairing. However, 
as will be seen below,
it is advantageous to keep the general 
form of the hamiltonian
(\ref{e:ham}).

One can find the exact ground state of (\ref{e:ham}) in the simple case
of one or two level systems. 
However, in the general case of a multilevel
system the dimension increases exponentially and therefore the
standard procedure is to 
use the generalized Bogolyubov transformation approach in the form 
\cite{Goswami}, where the quasiparticle operators are related to the
particle operators by
\begin{equation}
\left(
       \begin{array}{c}
                             c_{j1}^{\dagger} \\ 
                             c_{j2}^{\dagger} \\ 
                             c_{\overline{j}1} \\ 
                             c_{\overline{j}2} 
       \end{array}
                                                          \right) =
\left( 
        \begin{array}{cccc}
                             u_{11j} & u_{12j} & v_{11j} & v_{12j} \\ 
                             u_{21j} & u_{22j} & v_{21j} & v_{22j} \\ 
                    -v_{11j} & -v_{12j}^{*} & u_{11j} & u_{12j}^{*} \\ 
                        -v_{21j}^{*} & -v_{22j} & u_{21j}^{*} & u_{22j} 
        \end{array}
                                                         \right) 
\left( 
       \begin{array}{c}
                             a_{jp}^{\dagger} \\ 
                             a_{jn}^{\dagger} \\ 
                             a_{\overline{j}p} \\ 
                             a_{\overline{j}n} 
       \end{array}
                                                           \right) ~.
\label{e:bogol}
\end{equation}                                                           
Here $j$ denotes the full set of quantum numbers 
of a s.p. orbit, and the indeces ``1'' and ``2''  
are the quasiparticle analogs
of  $p$ or $n$, i.e. of the corresponding
isospin projections. The transformation amplitudes
$u_{ik,j}$ and $v_{ik, j}$ with $i \neq k$ describe 
the neutron-proton pairing.
They are, in general, complex.
We refer to \cite{Civitarese96}
and \cite{Goswami} for the unitarity conditions
which $u_{ik,j}$ and $v_{ik, j}$ have to obey, as well
as for the relation between the amplitudes and the gap
parameters $\Delta_p$, $\Delta_n$, and $\Delta_{np}$.

To find the ground state we minimize the quantity $H_0$, 
the expectation value of the hamiltonian in the quasiparticle 
vacuum, while simultaneously
obeying the unitarity conditions and the usual conservation 
(on average) of the number of neutrons and protons. 
(This procedure is equivalent to demanding
that the ``dangerous graph'' term $H_{20}$, which creates or 
annihilates a pair of quasiparticles, vanishes.) 
We use the Newton-Raphson method
\cite{recipes} and check, by comparing to the ``standard BCS'' solution 
for $G_{np} = 0$, that the ground state energy is lower 
than in the state without the neutron-proton pairing.
The procedure allows us to find at the same time the gain
in the ground state binding energy associated with the neutron-proton 
pairing.

\section{Isospin symmetric hamiltonian}

The hamiltonian (\ref{e:ham}) with $\epsilon _{jp} = \epsilon _{jn}$
and $G_{nn} = G_{pp} = G_{np}$ describes the 
isovector pairing, in which all
three kinds of pairs ($nn, ~ pp, $ and $np$ with $T=1$) 
are treated equally on the
interaction level. One expects then that in an 
even-even  nucleus with $N=Z$
the corresponding gap parameters should be the 
same for all three possible pairs.

In fact, in the exactly solvable manifestation of 
this hamiltonian, in which there
is only one s.p. state of degeneracy 2$\Omega$, this is indeed 
the case.  Defining the pair creation operator as
\begin{equation}
S_{t_z}^{\dagger} = \sum_{j,m>0} [ a_{jm}^{\dagger} 
a_{\overline{jm}}^{\dagger} ]_{t_z}^{T=1} ~,
\end{equation}
where $t_z$ is the corresponding isospin projection, the
quantity related to the pairing gap $\Delta_{t_z}$ is the ground state
expectation value ${\cal N}_{t_z} $ = 
$\langle S_{t_z}^{\dagger}S_{t_z} \rangle$. 
(We calculate the ``gap''  $\Delta_{t_z}$ from the expression
${\cal N}_{t_z} $ = $\Delta_{t_z}^2/G_{t_z}^2$ valid up to the terms
$1/\Omega$. This relation, however, fails for full shells.)
As shown in \cite{Engel1},
based on the earlier work on this $SO(5)$ model, one can obtain 
analytic expressions for ${\cal N}_{t_z} $. Indeed, when $N=Z$ and 
both are even, all three values of ${\cal N}_{t_z} $ are equal, and
when $N - Z$ increases the ${\cal N}_{0} $, 
and therefore also $\Delta_{np}$,
sharply decreases, while the other two ${\cal N}_{t_z = \pm 1} $
remain the same or
increase with $N-Z$. We expect that this behavior is ``generic'',
i.e. survives even in the case of more than one single
particle level.

The conjecture above is supported by the analysis of
the two level model \cite{Dussel}. If we restrict ourselves 
only to the states with seniority zero 
it is easy to construct the corresponding hamiltonian
matrix which has then very manageable dimensions
even for large $\Omega$. For completeness we give the
expressions for the corresponding matrix elements
in the Appendix. 

Unlike the exact solutions described above,
the generalized Bogolyubov transformation approach
gives a very different 
results in the isospin symmetric case.
It has been known for some time \cite{Covello} 
that in the one-level case
the approach leads to no $np$ pairing when $N > Z$. 
Our calculation shows
that this is true generally. There is, in fact, 
a {\it no mixing theorem}: The ground states
of even-even nuclei with $N > Z$ have vanishing 
$\Delta_{np}$ when isospin
symmetric hamiltonian is used. For $N = Z$ nuclei 
there is still no mixing. But in
that case there are two $degenerate$ minima of the energy; 
one with nonvanishing
$\Delta_n$ = $\Delta_p$ and $\Delta_{np} = 0$, and the other one with
$\Delta_n$ = $\Delta_p$ = 0 and $\Delta_{np} \neq 0$. 
(This conclusion was
also reached in \cite{Engel2} for the one-level model, 
and in \cite{Wyss} in the more general case.)

We see, therefore, that the generalized Bogolyubov 
transformation fails to describe
correctly the treatment of isovector pairing correlations. 
However, since we expect,
as stated above, that the effect of $np$ pairing decreases fast 
with increasing $N - Z$,
the standard BCS theory is still applicable for most heavier 
nuclei where $N-Z$ is relatively large. 

\section{Breaking the isospin symmetry}

Let us consider now what happens when the 
requirement of isospin symmetry is relaxed,
i.e., in the hamiltonian (\ref{e:ham}) one allows 
different coupling constants 
$G_{nn} \neq G_{pp} \neq G_{np}$, and possibly 
different single particle energies
for neutrons and protons.  It was shown already 
thirty years ago \cite{Goswami}
that such a hamiltonian, treated using the generalized Bogolyubov
transformation (\ref{e:bogol}), results in nonvanishing 
$\Delta_{np}$ in nuclei with $N \simeq Z$.

Hamiltonian which breaks isospin symmetry leads, 
naturally, to eigenstates that
do not have a definite value of isospin. 
Since the quasiparticle vacuum mixes 
states with different particle number, and therefore also with 
different isospin, even for 
an isospin conserving hamiltonian, it is perhaps 
worthwhile to explore effects
associated with such a more general situation.

It is straightforward to treat the hamiltonian (\ref{e:ham}) 
exactly in the one or two level models;
the corresponding hamiltonian matrix can be 
calculated using the formuae in
\cite{Hecht} (see also Appendix). The corresponding eigenstates 
are no longer characterized by
isospin $T$. Instead, all isospin values 
between $T _z \equiv T_{min} = (N - Z)/2$ 
and $T_{max} = (N + Z)/2$ contribute to the wave function. 
The ground state energy $E_{gs}$
of a one level system with 
$G_{np} \neq G_{pair} \equiv G_{nn} = G_{pp}$ 
decreases monotonically
with increasing $G_{np}$. However, the binding 
energy gain between a system
with no neutron-proton interaction (and therefore $\Delta_{np} = 0)$, 
and the system with pure isovector interaction 
(and $\Delta_{np} \neq 0$)
is only of the order $1/\Omega$,
\begin{equation}
\Delta E = E_{gs} (G_{np}/G_{pair} = 1) - E_{gs} 
(G_{np} = 0) = - G_{pair}Z/2 ~,
\end{equation}
compared to the leading term $-G_{pair}\Omega (N+Z)/2$.
Moreover, the exact wave function of the ground state
corresponding to the isovector pairing
hamiltonian with $G_{np}/G_{pair} = 1$ can be obtained from 
the ground state
of the isospin violating hamiltonian with {\it any} 
$G_{np}/G_{pair} \neq 1$
by simply projecting onto a state with isospin $T = T_{min}$. 
This is an exact statement which  follows from 
the uniqueness of the zero seniority state
with given $(N,Z,T)$.

In Fig. 1 we show the exact and BCS gap parameters for the one level 
system as a function of the ratio $G_{np}/G_{pair}$. 
The degeneracy $\Omega$ and the pairing strength $G_{pair}$ are
chosen in such a way  that they resemble the situation in finite
nuclei discussed later on. One can see that
the two methods give qualitatively similar results. 
They agree with each other
quite well, with the exception of the narrow region 
near the ``critical point''
of the BCS method (for the plotted case this point is at
$G_{np}/G_{pair} = 1.05$). As usual, the BCS method is characterized 
by the sharp phase transition while the exact method goes more smoothly
through the  ``critical point'', as it must for a finite system. 
Nevertheless, the basic similarity is apparent.

It is now clear that the failure of the BCS method to describe the 
neutron-proton pairing in the isospin symmetric case is not
a fundamental one. It is related to the abrupt phase transition
inherent in the BCS. The isospin symmetric value $G_{np} = G_{pair}$
is less than the critical value needed for the phase transition from
the pure like-particle pairing to the situation where both 
like-particle and neutron-proton pairs coexist. Since, as stated
earlier, the quasiparticle vacuum breaks isospin anyway, it should
not matter much that the isospin violation is also imposed on the
hamiltonian level. We have to choose, however, the proper value
of the coupling constant $G_{np}$. 

The natural way to fix the ratio $G_{np}/G_{pair}$ is in nuclei
with $N \simeq Z$ where one can use the arguments of isospin
symmetry to estimate the gap $\Delta_{np}$. We show in Fig. 2
again the comparison between the exact and BCS gaps in the
degenerate case, now as a function of $N-Z$. There are basic
similarities between the two situations, but the quantity
$\Delta_{np}$ decreases more rapidly with $N-Z$ in the
BCS case than in the exact case. We believe that this feature 
is related to the approximation involved in relating the gap
$\Delta_{t_z}$ to the ground state expectation 
value ${\cal N}_{t_z} $ in the exact case.
What is clearly visible in both cases, and intuitively
obvious, is the tendency of the $\Delta_{np}$
to decrease with increasing $N - Z$. This tendency have
been noted many times before, see e.g. \cite{Engel1,Civitarese96}.
In particular Ref. \cite{Civitarese96} have shown that in the BCS
approach for real nuclei, 
and with the ratio $G_{np}/G_{pair}$ fixed so that
at $N \simeq Z$ the gap $\Delta_{np}$ has reasonable value,
the effect of neutron-proton pairing disappears at $N - Z \geq 6$.

\section{Neutron-proton pairing and Congruence energy}

It has been known for some time, and reiterated recently in 
Refs. \cite{MS1,MS2} that nuclear masses can be well approximated
by a smooth Thomas-Fermi formula, shell correction terms, 
even-odd terms (i.e. standard like-particle pairing), and the 
congruence energy (or ``Wigner term'') which depends on $|N - Z|/A$. 
The least squares fit to the congruence energy gives \cite{MS2}
\begin{equation}
C(I) = -10 e^{-4.2 |I|} ~{\rm MeV}; ~ I = (N - Z)/A ~.
\label{e:cong}
\end{equation}

Satula and Wyss \cite{Wyss} suggested that the neutron-proton 
pairing interaction in the isospin $T=0$ (and therefore $J \neq 0$)
channel provides a microscopic origin of the Wigner term. In their
work \cite{Wyss} there is no neutron-proton $T=1$ pairing because
they use isospin symmetric hamiltonian and thus are missing
the neutron-proton $T = 1, J = 0$ pairing effects as explained above.
For the microscopic mechanism of the Wigner term in Ref. \cite{Wyss}
it is necessary that the strength of the $T=0$ pairing 
(called $x^{T=0}$ there) be above the
critical value $x^{T=0}$ = 1.

Here we consider only pairing interaction in time-reversed
orbits and thus the possible $T=0$ pairing is not present.
However, with the treatment of the $T=1$ 
neutron-proton pairing suggested
here we have an alternative (or additional) microscopic contribution
to the congruence energy, caused simply by the neutron-proton 
$T=1$ pairing. To see how this alternative
would work we compare in Fig. 3 the extra binding energies associated
with the $np$ $T=1$ pairing with the congruence energy $C(I)$ above.
(Since we are interested mainly in the $|N-Z|/A$ dependence we shifted
the $C(I)$ curve by 7 MeV.) As in Ref. \cite{Wyss} the extra binding
energy associated with pairing decreases with $|I|$ faster than the 
empirical congruence energy $C(|I|)$, at least for the degree of 
isospin symmetry breaking we have chosen. Obviously, the quantity
\begin{equation}
\delta B_{np} = E(G_{np} = 1.25 G_{pair}) - E(G_{np} = 0)
\label{e:diff}
\end{equation}
shown in Fig. 3 has qualitatively similar dependence on $(N - Z)$ as
the congruence energy, eq. (\ref{e:cong}). 
(The single particle level scheme end energies
of Ref. \cite{Goswami} were used in our calculation.)
It remains to be seen, however,
whether this contribution to the congruence energy 
is the only one (or at least the most important one).
Also, further investigation of the dependence of the gap $\Delta_{np}$
on the neutron excess is needed. It is possible that the BCS treatment
with the fixed ratio $G_{np}/G_{pair}$ exaggerates the slope of that 
dependence.

\section{Conclusions}

We have shown that treating pairing properties of a system
of interacting protons and neutrons with the help of the
generalized Bogolyubov transformation requires special care.
The corresponding system of equations allows, in principle, 
three different solutions. There is the trivial ``normal'' solution
with no pairing whatsoever. But there are also two competing
solutions with pairing correlations present. One, in which there
are no neutron-proton pairs, corresponds to a product state
with the neutron-neutron and proton-proton pairs not
communicating with each other. The other solution corresponds
to a system in which the like-particle and neutron-proton isovector
pairs coexist. When the generalized Bogolyubov 
transformation is used,
there is a sharp phase transition between
these two paired regimes, with the critical pairing strength
$G_{np}/G_{pair}$ somewhat larger than unity.
  
Thus, if one wants to describe the neutron-proton pairing using
the quasiparticle transformation method, one has to break isospin 
symmetry at the hamiltonian level. We propose to fix the unknown
degree of isospin breaking in such a way that the gap $\Delta_{np}$
in $N \simeq Z$ nuclei is reasonable, i.e. comparable to the
gaps $\Delta_{nn}$ and $\Delta_{pp}$. With this assignment
all traces of the isovector neutron-proton pairing disappear
for $N - Z \geq 6$ in real nuclei.

We conjecture that the isovector neutron-proton pairing represents one
of the main contributions to the congruence energy (Wigner term)
which describes the additional binding of 
nuclei with $(N - Z)/A \simeq 0$.
 
\acknowledgements

O.C. and M. R. are fellows of the  CONICET, Argentina.
O.C. acknowledges a grant of the J. S. Guggenheim Memorial Foundation
and thanks Caltech for hospitality.  The work of P.V.
was supported in part by the  U.\ S.\ Department 
of Energy under grant DE-FG03-88ER-40397.

\appendix
\section{}

For the exact diagonalization in the space of pairs with zero seniority 
we use the basis $|{\cal N},T,T_z \rangle$, where ${\cal N}$ is the 
total number
of pairs, $T$ is the isospin and $T_z$ its projection. 
The necessary matrix
elements of the pair creation and annihilation operators are
\begin{eqnarray}
\langle {\cal N}+1, T+1, T_z + t_z |  S_{t_z}^{\dagger} | 
{\cal N}, T, T_z \rangle  
&=& \nonumber \\
\frac{ \langle T T_z 1 t_z | T+1, T_z + t_z \rangle }{ \sqrt { 2T+3} }
[(T&+&1)(2\Omega-{\cal N}-T)(T+{\cal N}+3)/2]^{1/2},
\end{eqnarray}
\begin{eqnarray}
\langle {\cal N}+1, T-1, T_z + t_z |  S_{t_z}^{\dagger} | 
{\cal N}, T, T_z \rangle 
& = & \nonumber \\
\frac{ \langle T T_z 1 t_z | T-1, T_z + t_z \rangle }{ \sqrt { 2T-1} }
[T(2\Omega&+&1-{\cal N}+T)({\cal N} -T+2)/2]^{1/2},
\end{eqnarray}
\begin{eqnarray}
\langle {\cal N}-1, T+1, T_z - t_z |  S_{t_z} | 
{\cal N}, T, T_z \rangle  &=&
\nonumber \\
\frac{ \langle T T_z 1 -t_z | T+1, T_z - t_z \rangle }{ \sqrt { 2T+3} }
[(T&+&1)(2\Omega + 3 - {\cal N} + T)({\cal N} - T)/2]^{1/2},
\end{eqnarray}
\begin{eqnarray}
\langle {\cal N}-1, T-1, T_z - t_z |  S_{t_z} | 
{\cal N}, T, T_z \rangle & = &
\nonumber \\
\frac{ \langle T T_z 1 t_z | T-1, T_z - t_z \rangle }{ \sqrt { 2T-1} }
[T(2\Omega&+&2-{\cal N}-T)({\cal N} +T+1)/2]^{1/2}.
\end{eqnarray}

For the case of two levels the basis is 
$|{\cal N}_1,{\cal N}_2,T_1,T_2,T_{1z},T_{2z} \rangle$
with the obvious constraint 
${\cal N}_1 + {\cal N}_2 = {\cal N}$ (total number of pairs)
and a similar one for $T_z$. 
The hamiltonian matrix is easily constructed
from the expressions above, with terms 
diagonal in ${\cal N}_1, {\cal N}_2$ and terms
where ${\cal N}_1 \rightarrow {\cal N}_1 \pm 1$, and
 ${\cal N}_2 \rightarrow {\cal N}_2 \mp 1$. 
In addition there is a diagonal
shift of $2\epsilon {\cal N}_2$, where $\epsilon$ is 
the splitting of the single particle states.

\begin{figure}
\caption{ The pairing gaps for the one-level case with $\Omega$ = 11, 
$N$ = 6  and $Z$ = 4.
$G_{pair}$ = 0.242 MeV was used and the results
are plotted as a function of the ratio $G_{np}/G_{pair}$. 
The upper panel is for the
exact solution with gaps determined as described in the text. 
The lower panel is
for the BCS solution.}
\label{fig:deg1}
\end{figure}

\begin{figure}
\caption{ The pairing gaps for the one-level case with 
$\Omega$ = 11, as function
of the neutron excess $N - Z$. 
Both panels are for $Z$ = 4  while the neutron
number is varied; $G_{pair} = 16/(N+Z+56)$ MeV.
The exact solutions for $G_{np}/G_{pair}$ = 1.1 are in the upper panel.
The BCS solutions for $G_{np}/G_{pair}$ = 1.25 are in the lower panel.}
\label{fig:deg2}
\end{figure}

\begin{figure}
\caption{ The extra binding energy associated with 
the neutron-proton pairing,
eq.(\protect \ref{e:diff}) is depicted by the dashed lines
for the series of Ge (upper panel), 
Se (middle panel) and Kr (lower panel) isotopes. 
The full line, shown for comparison, 
represents the congruence energy, eq.(\protect \ref{e:cong}),
shifted by 7 MeV.
 }
\label{fig:cong}
\end{figure}

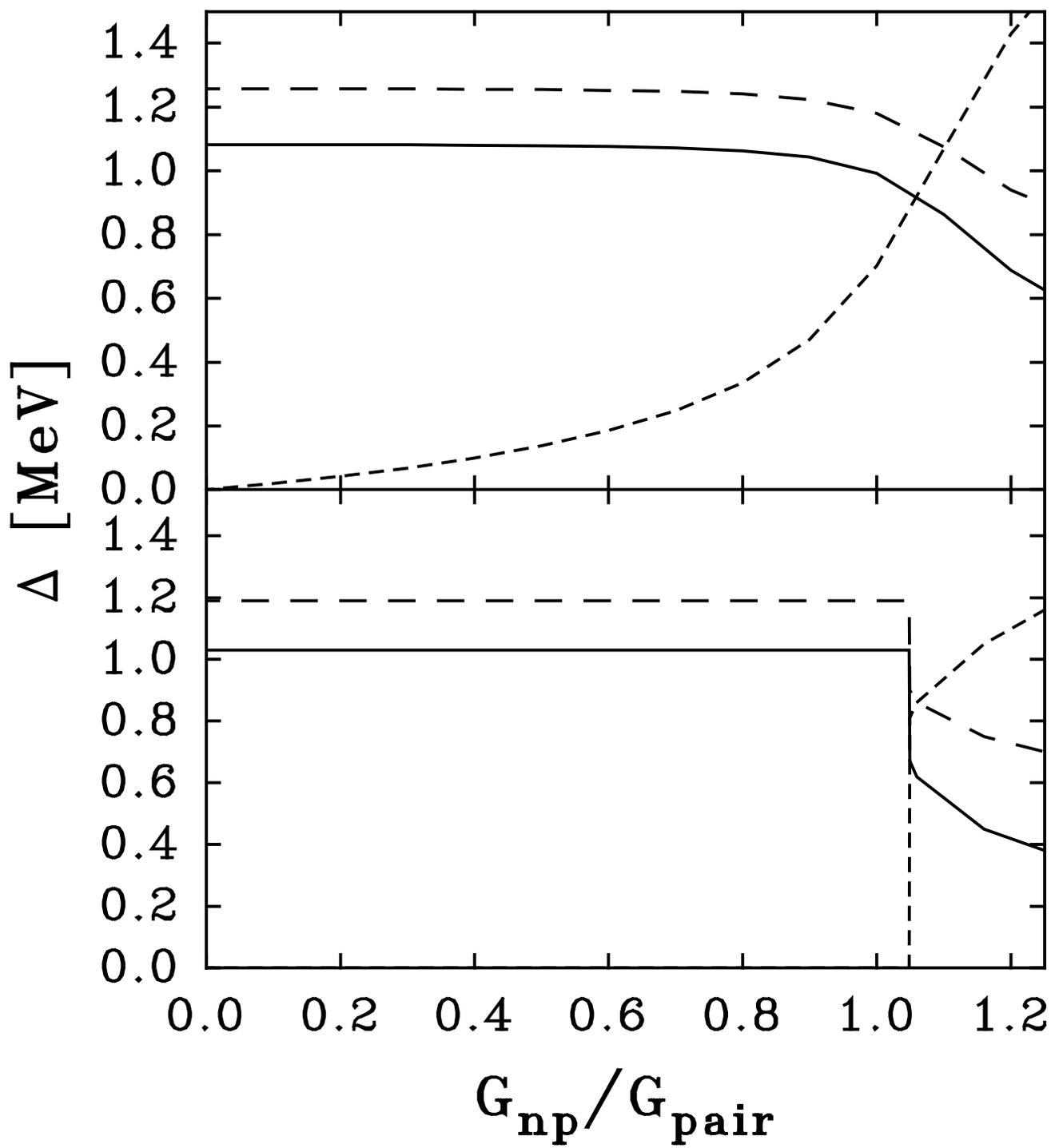
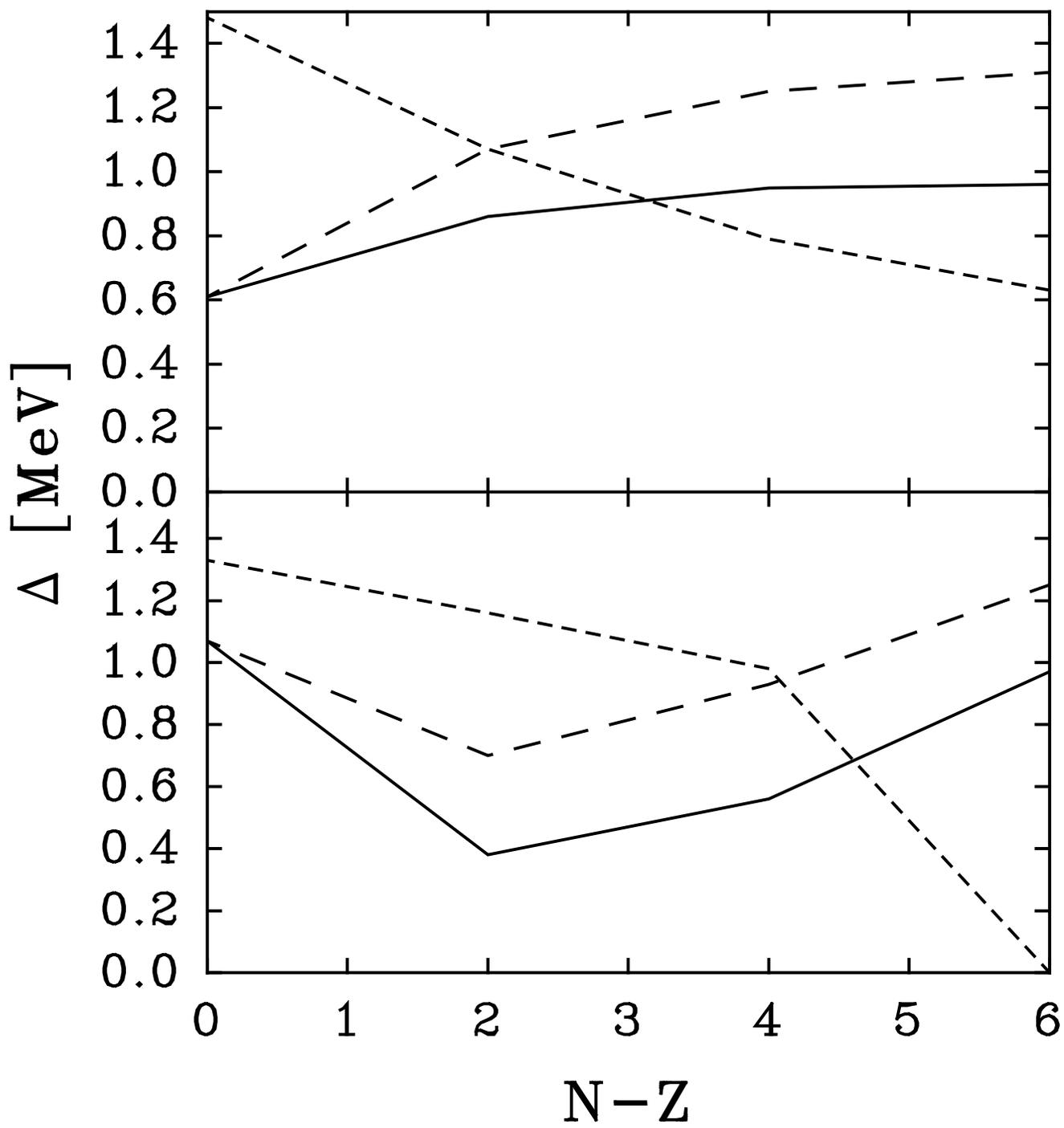
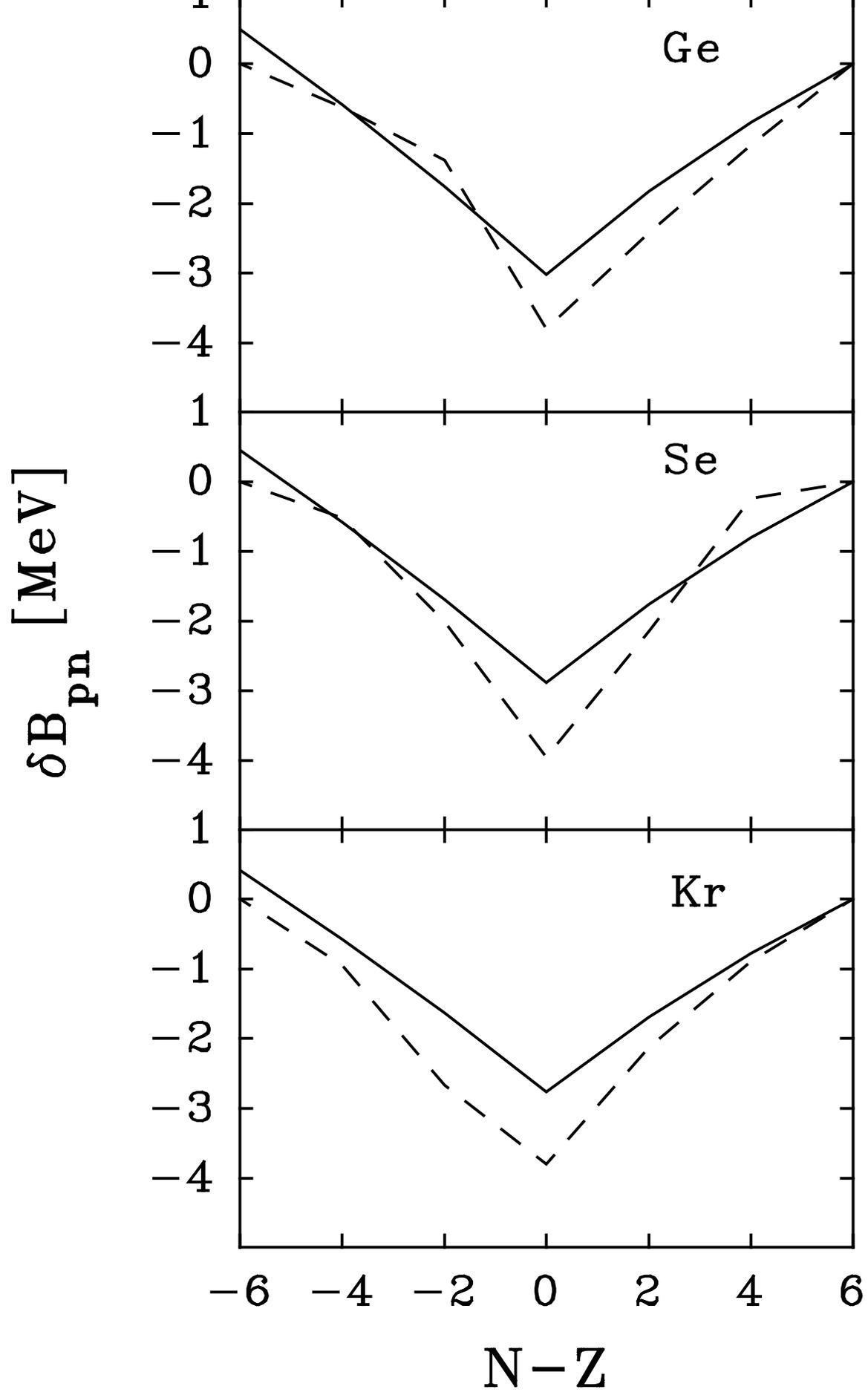
\end{document}